\documentclass[12pt,fleqn]{article}
\usepackage{times}
\usepackage[T1]{fontenc}
\normalfont
\usepackage{fourier} 
\usepackage{graphicx}
\usepackage{amssymb}
\usepackage{amsmath}
\usepackage{color}
\usepackage{cite}
\newcommand{\vev}[1]{{\langle #1 \rangle}}

\newcommand{\GeV}{\mbox{~GeV}}

\newcommand{\ie}{{\it i.e.}}

\begin{document}
\baselineskip 0.6cm
\renewcommand{\thefootnote}{\#\arabic{footnote}} 
\setcounter{footnote}{0}
%
\begin{titlepage}
\begin{center}

\begin{flushright}
\end{flushright}


{\Large \bf 
Initial condition for baryogenesis via neutrino oscillation
}\\

\vskip 1.2cm

{
Takehiko Asaka,$^1$
Shintaro Eijima,$^2$
Hiroyuki Ishida,$^3$
Kosuke Minogawa,$^4$
and 
Tomoya Yoshii$^4$
}

\vskip 0.4cm

$^1${\em \small
  Department of Physics, Niigata University, Niigata 950-2181, Japan
}

$^2${\em \small
 Institute of Physics, Laboratory for Particle Physics and Cosmology, \'Ecole Polytechnique F\'ed\'erale de Lausanne, CH-1015 Lausanne, Switzerland
}

$^3${\em \small
  Physics Division, National Center for Theoretical Sciences, Hsinchu, 30013 Taiwan
}

$^4${\em \small
Graduate School of Science and Technology, Niigata University, Niigata 950-2181, Japan}

\vskip 0.2cm

(April 10, 2017)

\vskip 2cm

\begin{abstract}
    We consider a baryogenesis scenario via the oscillation of
    right-handed neutrinos with Majorana masses of the order of GeV, which are also 
    responsible for neutrino masses by the seesaw mechanism.  
    We study how the initial condition  alters the
    prediction of the present baryon asymmetry by this mechanism.
    It is usually assumed that the abundance
    of right-handed neutrinos is zero 
    after the reheating of the inflationary universe and they are 
    produced in scattering processes by the renomalizable 
    Yukawa interaction.  However, the higher-dimensional operator 
    with right-handed neutrinos may provide an additional production 
    which is most effective at the reheating epoch.
    It is shown that such an initial abundance of right-handed neutrinos 
    can significantly modify the prediction 
    when the strong washout of the asymmetry is absent. 
    This leads to the parameter space of the model for the successful baryogenesis being enlarged. 
\end{abstract}
\end{center}
\end{titlepage}

\section{Introduction}
The baryon asymmetry of the universe (BAU) is one of the most mysterious
numbers in particle physics and cosmology.
The present baryon-to-entropy ratio, $Y_B = n_B/s$ with the baryon 
number density $n_B$ and the entropy density $s$, 
is observed as~\cite{Ade:2015xua}
\begin{align}
  \label{eq:YBobs}
  Y_B^{\rm obs}
  = (8.677 \pm 0.054) \times 10^{-11} \,.
\end{align}
When the universe experiences the inflation, this asymmetry must be
generated after the exponentially expanding period, 
more precisely, during or after the reheating process, which gives an initial condition of the hot big-bang universe. 
Such a mechanism, called as baryogenesis, requires three conditions~\cite{Sakharov:1967dj}.  
Unfortunately, the standard model (SM)
cannot satisfy some of these conditions.  
Especially, since the Higgs boson mass is measured as
125~GeV, the electroweak phase transition is a smooth crossover 
(See, for example, the recent analysis~\cite{DOnofrio:2015gop}
and references therein.)
and then there is no way to lead a deviation from thermal equilibrium which is one of the above conditions.
Thus, it is important to clarify new physics 
in order to explain the BAU between the inflation and the big-bang nucleosynthesis.

The baryogenesis scenarios with right-handed neutrinos are attractive
since they can simultaneously account for tiny neutrino masses through
the seesaw mechanism~
\cite{Minkowski:1977sc,Yanagida:1979as,Yanagida:1980xy,GellMann:1980vs,Ramond:1979,Glashow:1979,Mohapatra:1979ia}.
In the (canonical) leptogenesis scenario~\cite{Fukugita:1986hr}, the
out-of-equilibrium decays of right-handed neutrinos can generate the
lepton asymmetry which can be partially converted to the baryon
asymmetry by the rapid sphaleron process for the temperatures
$T>{\cal O}(10^2)$~GeV~\cite{Kuzmin:1985mm}.  The observed value
(\ref{eq:YBobs}) can be explained when hierarchical right-handed
neutrinos are heavier than ${\cal O}(10^9)$ GeV.
(See, for example, Refs.~\cite{Giudice:2003jh,Buchmuller:2005eh,Davidson:2008bu}.)

On the other hand, the flavor oscillation between right-handed
neutrinos can also be responsible for the BAU~\cite{Akhmedov:1998qx,Asaka:2005pn}.  
In this case the observed value (\ref{eq:YBobs}) can be generated even when the Majorana masses
are below the electroweak scale as long as the masses are quasidegenerate.
One of the attractive models which realizes this baryogenesis 
is the neutrino minimal standard model ($\nu$MSM)~[18,17]. 
In this scenario, 
two heavier ones among three right-handed neutrinos are responsible for baryogenesis 
and the remaining lightest one is a candidate of dark matter.

Since these right-handed neutrinos have tiny Yukawa coupling constants to realize the seesaw mechanism 
as well as to avoid the cosmological constraints on dark matter, 
the production rates by the scattering of the SM particles at the reheating epoch is negligible and then 
one can consider that there is essentially no right-handed neutrino in the beginning of the hot big-bang universe. 
However, unknown physics at high energy scale could 
produce right-handed neutrinos through the interactions suppressed by the cutoff scale of the model. 
One interesting possibility is the dimension-five interaction with right-handed neutrinos~\cite{Bezrukov:2008ut}.

Although such an additional production is possible to affect the generation of the BAU, 
it has been shown in Ref.~\cite{Bezrukov:2008ut} that the effect is negligible 
if the dark matter abundance which is equally produced as the heavier ones does not overclose the universe. 
Here, we revisit this problem in several situations for the baryogenesis via neutrino oscillation. 
The important physical quantities for this scenario are the starting temperature of the right-handed neutrino oscillation $T_{\rm osc}$ 
which quantifies the oscillation effect to baryogenesis and 
the overall magnitude of the neutrino Yukawa coupling constants which quantifies the washout effect of the asymmetries. 
We will then consider several choices for these quantities leading to different situations of this baryogenesis. 

As we will show below, 
when the neutrino Yukawa coupling constants are large enough, 
the initial abundance of the right-handed neutrinos does not affect on the produced BAU due to the strong washout effects. 
This result is consistent with the previous analysis. 
On the other hand, in the weak washout region, 
the situation is drastically changed and 
the impact on baryogenesis can be classified by $T_{\rm osc}$ and the initial condition of the matrices of densities. 
The off-diagonal element of the initial matrices of densities can significantly affect to the produced BAU when it is sufficiently large. 
As $T_{\rm osc}$ decreasing, the effects on the produced BAU from all elements of the initial matrices of densities are significant. 
In addition, the CP violating effects in the initial values of the matrices of densities can offer an additional source of the BAU. 
These conclusions are different from the previous one. 

This paper is organized as follows. 
We will present our framework in the next section. 
The impacts on the dark matter physics and baryogenesis will be discussed in Secs.~\ref{Sec:DM} and \ref{Sec:BAU}. 
We will conclude our paper in Sec.~\ref{Sec:conclusion}. 
In addition, we review the production of right-handed neutrinos via higher-dimensional operator in App.~\ref{sec:ap1} 
and the parametrization of the neutrino Yukawa coupling constants in App.~\ref{sec:ap_Yukawa}. 
Finally, we present the evolution of the yield of the BAU in various situations in App.~\ref{Sec:app3}.

\section{The framework}\label{Sec:framework}
We consider the $\nu$MSM,  
an extension of the SM by three right-handed neutrinos $\nu_{R I}$ ($I=1,2,3$).
The Lagrangian is given by
\begin{align}
  {\cal L}_{\nu{\rm MSM}}
  =
  {\cal L}_{\rm SM}
  + i \, \overline{\nu_{RI}} \gamma^\mu \partial_\mu \nu_{RI}
  -
  \left(
  F_{\alpha I} \, \overline{\ell_{\alpha}} \, \Phi \, \nu_{RI}
  +
  \frac{M_I}{2} \, \overline{\nu_{RI}^c} \, \nu_{RI}
  +
  H.c.
  \right) \,,\label{Eq:Lag}
\end{align}
where ${\cal L}_{\rm SM}$ is the SM Lagrangian, and $\Phi$ and
$\ell_\alpha$ ($\alpha = e, \mu, \tau$) are the weak doublets for
Higgs and left-handed leptons, respectively.
$F$ is the matrix of Yukawa coupling constants
and $M_I$ is the Majorana mass of right-handed neutrino.
Here and hereafter we shall work in the basis where
the mass matrix of charged leptons and the Majorana mass matrix $M_M = {\rm diag}(M_1\,,M_2\,,M_3)$ are both diagonal.

We require the hierarchy between the Majorana mass $M_I$ and the Dirac
mass $[M_D]_{\alpha I} = F_{\alpha I} \langle \Phi \rangle$
($\langle \Phi \rangle = 174~{\rm GeV}$ is the Higgs vacuum expectation value),
$|[M_D]_{\alpha I}| \ll M_I$, to realize the seesaw mechanism.  In
this case, the mass eigenstates are active neutrinos $\nu_i$
($i=1,2,3$) and heavy neutral leptons (HNLs) $N_I$.  The left-handed
neutrino is then written as
\begin{align}
  \nu_{L \alpha } = U_{\alpha i} \, \nu_i + \Theta_{\alpha I} \, N_I^c \,,
\end{align}
where $U$ is the Pontecorvo-Maki-Nakagawa-Sakata (PMNS) mixing matrix for active neutrinos~\cite{PMNS} 
and $\Theta = M_D/M_M$ is that for HNLs.  The
masses of active neutrinos $m_i$ are determined from
\begin{align}
  [M_\nu]_{\alpha \beta} = - [M_D]_{\alpha I} [M_D]_{\beta I} M_I^{-1} \,,
  ~~~~
  U^\dagger M_\nu U^\ast = \mbox{diag}( m_1 \,,~ m_2 \,,~ m_3 ) \,.
\end{align}
On the other hand, $N_I \simeq \nu_{RI}$ and the masses of HNLs are given by $M_I$.

In the $\nu$MSM, the lightest HNL $N_1$ is a candidate for dark matter
and the heavier ones $N_2$ and $N_3$ realize the seesaw mechanism of 
active neutrino masses and baryogenesis via neutrino oscillation for the BAU.  
Since the dark matter can radiatively decay into pairs of active neutrino and photon, 
the mixing angle is highly constrained from x-ray observations depending on its mass~\cite{Boyarsky:2006fg}. 
According to such severe constraints, 
the dark matter abundance is difficult to be explained by the nonresonant production~\cite{Dodelson:1993je} 
together with the constraint from the structure formation~\cite{Boyarsky:2008xj}. 
On the other hand,  
the smaller mixing region is possible to explain the dark matter density by invoking the resonant production~\cite{Shi:1998km} 
in the presence of large lepton asymmetry at the production period. 
In any cases, since the Yukawa coupling constants of $N_1$ must be highly suppressed, 
it plays substantially no role for the seesaw mechanism and baryogenesis. 
As a result, the Yukawa coupling constants of $N_{2\,,3}$ should be chosen 
in order to explain the tiny active neutrino masses. (See App.~B.) 
Thanks to the smallness of the Majorana masses, their coupling constants are found to be small.

The production of these right-handed neutrinos at high temperature region ($T \gg 10^{2}~{\rm GeV}$) is ineffective 
as long as it is governed by the Yukawa interaction. 
This is because the required coupling constants are suppressed 
and the production rate being proportional to $T$ is much 
slower than the expansion rate proportional to $T^2$. 
Consequently, it is considered that the abundance of right-handed neutrinos is essentially zero after the inflation. 
This is the conservative initial condition for considering dark matter and baryogenesis in the $\nu$MSM.

In this analysis we study the effects on physics of HNLs
by a higher-dimensional operator
\begin{align}
  \label{eq:HD}
  {\cal L}_{\rm HD}
  = \frac{A_{IJ}}{2 \Lambda} \, \Phi^\dagger \Phi \,
  \overline{\nu_{RI}^c} \, \nu_{RJ} + H.c. \,,
\end{align}
where $\Lambda$ is a cutoff scale and $A$ is a coupling matrix.
It should be noted that $A$ is not a diagonal matrix in general, which
is crucial for baryogenesis as will be shown later.  We assume that
all the elements $A_{IJ}$ are of the same order of magnitude ({\it i.e.,} of the order of unity) 
for simplicity.  The impact of the operator (\ref{eq:HD}) has already
been discussed in Ref.~\cite{Bezrukov:2008ut}, especially in the
context of the Higgs inflation.  Here, we revisit this problem paying
a special attention to the impact on baryogenesis.  In order to make a
more general discussion we take the reheating temperature $T_R$ as
a free parameter without specifying the inflation model.

One of the impacts by the operator (\ref{eq:HD}) is a correction to
the Majorana masses for right-handed neutrinos 
\begin{align}
  [\delta M_M]_{IJ}
  =
  - A_{IJ} \frac{\langle \Phi \rangle^2}{\Lambda}
  = 
  - 1.3 \times 10^{-5}~\mbox{eV} \, 
  A_{IJ} \,
  \left( \frac{M_P}{\Lambda} \right) \,,
\end{align}
where $M_P \simeq 2.4 \times 10^{18}$~GeV is the reduced Planck mass.  
It is seen that such corrections are significantly smaller than the masses of HNLs,
namely much smaller than the typical mass ${\cal O}(10)$~keV of the sterile neutrino dark matter,
and then they give only negligible impact on the masses of active neutrinos and HNLs.
It should, however, be noted that such a correction
might be important for the mass difference between $N_2$ and $N_3$, which is a key parameter for the baryogenesis via neutrino oscillation.
In this case the mass difference used to estimate the BAU
should be considered as the one given by $M_M + \delta M_M$.

Another consequence of the operator~(\ref{eq:HD}) is the additional
production of right-handed neutrinos by scatterings of Higgs particles
at the reheating epoch~\cite{Bezrukov:2008ut}.  
Note that the production rate due to the operator scales as
$T^3$ and then a sizable amount of $\nu_R$ may be generated at the reheating epoch.  

Let us consider the scattering process by the operator (\ref{eq:HD})
\begin{align}
  \phi_a + \phi_a \to \nu_{R_I} + \nu_{R_J} \,,\label{Eq:scatter}
\end{align}
where four real scalar fields of Higgs doublet are denoted by 
$\phi_a$ ($a=1,2,3,4$).  
We introduce the matrices of densities of the right-handed neutrino states with the positive and negative helicities 
\begin{align}
  [\rho_N (\vec k)]_{IJ} 
  = 
    \frac{1}{V}
    \mbox{Tr}\left[ \, \hat \rho(t) \, 
    \hat a^\dagger_J (\vec k\,,~ +) \, \hat a_I (\vec k\,,~+) \, \right] \,,\hspace{0.5cm}
  [\rho_{\overline N} (\vec k)]_{IJ} 
  = 
    \frac{1}{V}
  \mbox{Tr}\left[ \, \hat \rho(t) \, 
  \hat a^\dagger_J (\vec k \,,~-) \, 
    \hat a_I (\vec k \,,~-) \, \right] \,,
\end{align}
where $V$ is the space volume and $\hat \rho(t)$ is the density matrix operator.  
The annihilation and creation operators of $\nu_{R_I}$ with helicity $h$ and momentum $\vec k$
are denoted by $\hat a_I (\vec k\,, h)$ and
$\hat a_I^\dagger (\vec k\,,h)$, respectively.
As explained in App.~\ref{sec:ap1}, the contributions from the production process~(\ref{Eq:scatter}) 
are estimated as
\begin{align}
  \label{eq:rN_I}
  [r_N]_{IJ} = 9.4 \times 10^{-4} 
  \frac{M_P \, T_R}{\Lambda^2}  
  \left[ A^\dagger A \right]_{IJ} \,,~~~~~
  [r_{\overline N}]_{IJ} = 9.4 \times 10^{-4} 
  \frac{M_P \, T_R}{\Lambda^2}  
  \left[ A^T A^\ast \right]_{IJ} \,.
\end{align}
Here $r_N$ and $r_{\overline N}$ are the matrices of densities for right-handed neutrinos 
which are normalized by the equilibrium distribution function 
$f^{\rm eq} = ( e^{|\vec k|/T}  + 1 )^{-1}$.
Notice that we have neglected the masses of right-handed neutrinos
since $T_R \gg M_I$. 

The initial values of the matrices of densities after the reheating,
$[r_N]^I$ and $[r_{\overline N}]^I$, are then given by Eq.~(\ref{eq:rN_I}).
It should be noted that any asymmetry between $\nu_{R}$ and $\overline \nu_{R}$ is not generated
at the leading order since $[r_{\overline N}]^I = ([r_N]^I)^\ast$ as shown
in Eq.~(\ref{eq:rN_I}).
From now on we shall discuss the impacts of such initial abundances
on dark matter and baryogenesis by $\nu_R$.

\section{Impact on dark matter}\label{Sec:DM}
Now let us study the first topic of this article, the impact of the
operator (\ref{eq:HD}) on dark matter physics.  In the $\nu$MSM the
lightest HNL $N_1$ is a dark matter candidate.  In the conventional
scenario $N_1$ is produced by the scattering processes through the
weak interactions with the mixing
$\Theta_{\alpha 1}$~\cite{Dodelson:1993je},
and the dominant production occurs at the temperature $T_\ast \sim 100$~MeV
for $M_1 \sim $ keV. 
The present abundance of dark matter is given by~\cite{Ade:2015xua}
\begin{align}
  \label{eq:OMdm}
  \Omega_{\rm dm} h^2 = 0.1188 \pm 0.0010 \,,
\end{align}
where $\Omega_{\rm dm}$ is the present density parameter of dark 
matter component and $h$ is the present Hubble parameter in unit of 100~km/sec/Mpc.  This abundance can be explained by requiring the
mixing elements as~\cite{Asaka:2006nq}
\begin{align}
  \label{eq:THsqDM}
  \sum_{\alpha} |\Theta_{\alpha 1}|^2 \simeq 8 \times 10^{-8}
  \left( \frac{M_1}{1~\mbox{keV}} \right)^{-2} \,.
\end{align}
Such a dark matter candidate is constrained
from various cosmological observations.  
From the cosmic x-ray observations, the radiative decay 
$N_1 \to \nu + \gamma$ is severely restricted and the strong upper 
bound on $\sum_\alpha |\Theta_{\alpha 1}|^2$ is obtained~\cite{Boyarsky:2006fg}.
In addition, $N_1$ with keV mass, which plays a role of warm dark matter, receives the constraint from the cosmic structure
and the lower bound on $M_1$ is obtained~\cite{Boyarsky:2008xj}.
The present status of this dark matter scenario is difficult to realize within the simplest thermal history.
See the discussion in Ref.~\cite{Boyarsky:2009ix,Adhikari:2016bei}.

On the other hand, when there exists a sizable lepton asymmetry
around $T \simeq T_\ast$, the resonant production of $N_1$
occurs~\cite{Shi:1998km}.  In this case the mixing elements much
smaller than Eq.~(\ref{eq:THsqDM}) can account for the dark matter
abundance~(\ref{eq:OMdm}) and the severe x-ray bound can be avoided.
See the analysis in Ref.~\cite{Laine:2008pg}.  Interestingly, as shown
in Refs.~\cite{Canetti:2012vf,Canetti:2012kh}, the required lepton
asymmetry can be originated in the dynamics of $N_2$ and $N_3$ without
introducing any new physics.

\begin{figure}[t]
  \centerline{
  \hspace{-7mm}
  \includegraphics[width=7.5cm]{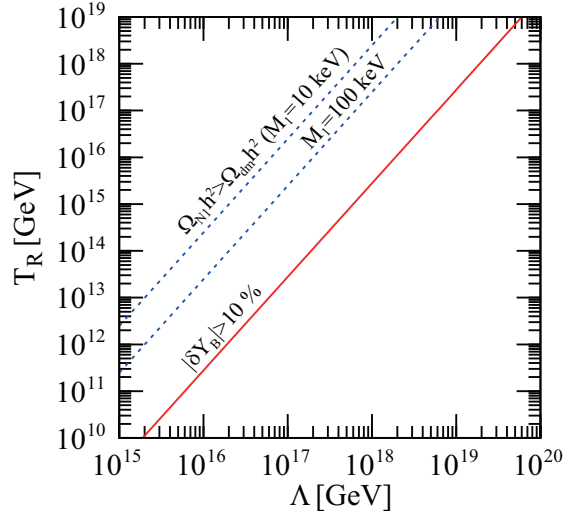}
  }%
  \caption{
    Upper bounds on the reheating temperature $T_R$ in terms of the cutoff scale $\Lambda$. 
    The bounds from the dark matter abundance $\Omega_{N_1}h^2 < \Omega_{\rm dm} h^2$ 
    for $M_1$=10~keV and 100~keV are shown by the blue dashed lines.
    The bound from the BAU $|\delta Y_B| < 10$~\% when $\Delta M = 10^{-13}$ GeV 
    and $X_\omega=1$ is also shown by the red solid line.}
  \label{fig:FIG_DM_2}
\end{figure}

Now we turn to discuss the impact of the higher-dimensional operator (\ref{eq:HD}).  As pointed out in Ref.~\cite{Bezrukov:2008ut}, 
enough numbers of $N_1$ can be generated at the reheating epoch.
We can estimate the density parameter of $N_1$ 
from the initial values of $[r_N]_{11}^I$ and $[r_{\overline N}]_{11}^I$ as
\begin{eqnarray}
  \label{eq:OMN1}
  \Omega_{N_1} h^2 = 
  0.11 
  \left(
  \frac{\left( [r_N]_{11}^I + [r_{\overline N}]_{11}^I \right)}{10^{-2}}
  \right)
  \left( \frac{M_1}{10~\mbox{keV}} \right) \,.
\end{eqnarray}
Note that $[r_N]_{11}^I =[r_{\overline N}]_{11}^I$.
%
\begin{figure}[t]
  \centerline{
  \includegraphics[width=10cm]{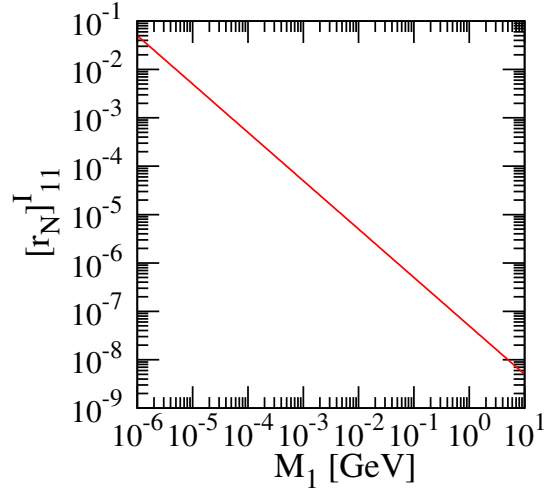}
  }%
  \vspace{-2ex}
  \caption{
    Initial value of $[r_N]^I_{11}$
    required for the correct dark matter density
    $\Omega_{N_1} = \Omega_{\rm dm}$.
    The region above the red line is excluded because of the overclosure of the universe. 
  }
  \label{fig:FIG_DM}
\end{figure}
It is thus found that
the initial value of $[r_N]^I_{11}$ must be
\begin{eqnarray}
  \label{eq:UBrN11}
  [r_N]_{11}^I < 5.6 \times 10^{-3} \,
  \left( \frac{10~\mbox{keV}}{M_1} \right)
  \,,
\end{eqnarray}
in order that the abundance $\Omega_{N_1}$ in Eq.~(\ref{eq:OMN1})
does not exceed the observed value (\ref{eq:OMdm}),
which is shown in Fig.~\ref{fig:FIG_DM}.
This bound can be translated into the upper bound on $T_R$, but it is a mild bound.  
For instance, when $\Lambda = M_P$, $A={\cal O}(1)$, and $M_1 > \mathcal{O} (10^2)~{\rm keV}$, 
the condition can be satisfied even if $T_R=M_P$. 
See Fig.~\ref{fig:FIG_DM_2}.

When the mixing of $N_1$ is extremely smaller than the value (\ref{eq:THsqDM})
and the initial values of the matrices of densities
take the value as shown in Fig.~\ref{fig:FIG_DM},
the alternative scenario of dark matter is possible~\cite{Bezrukov:2008ut}.%
\footnote{
It is possible to realize a mixed dark matter scenario
where both $N_1$'s produced at the reheating and $T \simeq $100 MeV
periods contribute the present dark matter density comparably.}  
For example, when $\Lambda = M_P$, $T_R=10^{13}$ GeV, and 
$A={\cal O}(1)$, $N_1$ with mass $M_1 = 4.9$ GeV becomes
a viable candidate for dark matter.
One might worry about the stability of such a heavy particle.
It should be noted, however, that 
we can switch off the Yukawa coupling for $N_1$
(\ie, $F_{\alpha 1}$=0) since the operator (\ref{eq:HD}) 
is responsible for the production.
In this case, the discrete symmetry
$Z_2$ associated with $N_1$ arises, under which only
$N_1 = \nu_{R1}$ changes as $N_1 \to - N_1$.  This symmetry,
in which the operator (\ref{eq:HD}) is invariant, ensures
the complete stability of $N_1$.  On the other hand,
the tests of such a dark matter candidate become difficult
since the mixing of $N_1$ disappears.
The test in the cosmic x-ray background~\cite{Adhikari:2016bei} is 
impossible because $N_1 \to \nu \gamma$ is forbidden
and the test at KATRIN~\cite{Mertens:2014osa,Mertens:2014nha} is also 
impossible because $N_1$ cannot be produced in tritium 
beta decay.

Finally, we would like to comment on the impact on the DM scenario specific to the $\nu$MSM.  
As we mentioned before, the heavier HNLs $N_{2,3}$ have to generate the lepton asymmetry via oscillation mechanism 
not only for the baryon asymmetry 
but also for the resonant production of the DM~\cite{Shi:1998km} after sphaleron freeze-out~\cite{Canetti:2012kh}. 
Below the sphaleron freeze-out temperature $N_{2,3}$ get in thermal equilibrium 
and decouple just before the lepton asymmetry generation for the resonant DM production.  
This means that the initial values of the matrices of densities
are irrelevant for the generation of the lepton asymmetry mechanism for the DM 
since the initial information is lost through the above thermalization.
However, the higher-dimensional operator gives the impact differently.
As we will show below, thanks to the initial value $[ r_N ]_{IJ}^I$ ($I,J=2,3$), 
the model parameter space to realize the correct amount of the BAU becomes wider than the case of $[ r_N ]_{IJ}^I = 0$.
In this case the late-time lepton asymmetry production discussed in~\cite{Canetti:2012kh} 
could be achieved in the relaxed parameter space.
This enlarges the possible space of the resonant DM production.
However, the quantitative analysis of the resonant DM production is
beyond the scope at the moment.

\section{Impact on baryogenesis}\label{Sec:BAU}
Next, we turn to discuss the impact of the initial abundance
(\ref{eq:rN_I}) on baryogenesis.
In the considering model, the sufficient amount of the BAU
can be generated by the mechanism via neutrino oscillation, namely 
by the oscillation between the quasidegenerate $N_2$ and $N_3$~\cite{Akhmedov:1998qx,Asaka:2005pn}.  
We then write $M_2 = M_N - \Delta M/2$ and $M_3 = M_N + \Delta M/2$
with $M_N \gg \Delta M$.
Such an impact has also been studied in Ref.~\cite{Bezrukov:2008ut}.
It has been shown that, when all the initial values of the matrices of densities are
sufficiently small as shown in Eq.~(\ref{eq:UBrN11}) in order to avoid
the overproduction of dark matter particle, the yield of the BAU does not
change significantly.  We shall revisit this problem paying a special
attention to the initial value of the off-diagonal element
$[r_N]_{23}^I$.  Such an element can be generated at the reheating
epoch because the coupling matrix $A$ is not diagonal in general and
the basis of right-handed neutrinos at high temperature regions is
different from the vacuum one. See the discussion in App.~\ref{sec:ap1}.
As we will show below,
the initial value of $[r_N]_{IJ}^I$ can alter the final BAU significantly 
even if they satisfy the condition~(\ref{eq:UBrN11}),
which should be contrasted to the previous results in Ref.~\cite{Bezrukov:2008ut}. 

We estimate the yield of the BAU $Y_B$ by solving the coupled equations
for the matrices of densities $[\rho_N]$ and $[\rho_{\overline N}]$ for
right-handed neutrinos with positive helicity and their antiparticles
with negative helicity, respectively as well as chemical potentials
for (left-handed) leptons.  When the neutrino Yukawa coupling
constants are sufficiently small, there are three conserved charges
$X_\alpha = B/3 - L_\alpha$, where $B$ is the baryon number
and $L_\alpha$ is the lepton number of a specific flavor $\alpha$.
The chemical potential of $X_\alpha$ (which is divided by the
temperature and is dimensionless) is denoted by $\mu_{X_\alpha}$,
while the chemical potential of $L_\alpha$ is $\mu_\alpha$, and they
satisfy
\begin{align}
  \mu_{\alpha} &= - \sum_{\beta} C_{\alpha\beta} \, \mu_{X_{\beta}}  \,,
\end{align}
where
\begin{eqnarray}
C_{\alpha\beta} = \frac{1}{711}\left(
	\begin{array}{ccc}
		221 & -16 & -16 \\
		-16 & 221 & -16 \\
		-16 & -16 & 221
	\end{array}
	\right) \,.
\end{eqnarray}

The equations for $[r_N]$, $[r_{\overline N}]$ and $\mu_{X_\alpha}$ 
have been discussed in Refs.~\cite{Akhmedov:1998qx,Asaka:2005pn,Canetti:2012vf,Canetti:2012kh,Asaka:2011wq,Shaposhnikov:2008pf,Canetti:2010aw,Khoze:2013oga,Canetti:2014dka,Shuve:2014zua,Garbrecht:2014bfa,Abada:2015rta,Hernandez:2015wna,Kartavtsev:2015vto,Drewes:2016gmt}.
In this analysis, we take into account the interaction rates induced by
the top Yukawa and gauge interactions and also their temperature dependence 
by renormalization group equations effect.\footnote{
The effects of the lepton number violation~\cite{Eijima:2017anv,Ghiglieri:2017gjz} 
and the CP violating Higgs decays~\cite{Hambye:2017elz,Hambye:2016sby} have been discussed in recent years, 
which are missed in the present analysis,
but such effects are expected not to change the final results of this paper qualitatively.
}
 See the details, for example, in Ref.~\cite{Hernandez:2016kel}. 
\begin{align}
  \label{eq:KE1}
  \frac{dr_N}{dt} 
  & = 
    -i \, \bigl[ \langle H \rangle \,,~ r_N \bigr] 
    - \frac{\langle \gamma^{(0)}_N \rangle}{2} \,
    \bigl\{ F^{\dag}F \,,~ r_N-1 \bigr\} 
    + \langle \gamma^{(1)}_N \rangle F^{\dag}\mu F  
    - \frac{\langle \gamma^{(2)}_N \rangle}{2} \,
    \bigl\{ F^{\dag} \mu F \,,~ r_N \bigr\}  \,,
  \\
  \label{eq:KE2}
  \frac{dr_{\overline{N}}}{dt} 
  & = 
    -i \, \bigl[ \langle H^\ast \rangle \,,~ r_{\overline{N}} \bigr] 
    - \frac{\langle \gamma^{(0)}_N \rangle}{2} \,
    \bigl\{ F^{T}F^{*} \,,~ r_{\overline{N}}-1 \bigr\}
    - \langle \gamma^{(1)}_N \rangle F^{T} \mu F^{*} 
    + \frac{\langle \gamma^{(2)}_N \rangle}{2} \,
    \bigl\{ F^{T} \mu F^{*} \,,~ r_{\overline{N}} \bigr\}  \,,
  \\
  \label{eq:KE3}
  \frac{d\mu_{X_{\alpha}}}{dt} 
  & = 
    - \frac{9 \, \zeta (3)}{ \pi^2 } \,
    \Biggl\{ \frac{\langle \gamma_{N}^{(0)} \rangle}{2}(Fr_{N}F^{\dag} 
    - F^{*}r_{\overline{N}}F^{T})_{\alpha \alpha}
    \nonumber \\
  &\hspace{10ex}
     - \sum_{\beta} C_{\alpha\beta} \, \mu_{X_{\beta}} 
    \Bigl[ \frac{\langle \gamma_{N}^{(2)} \rangle}{2}(Fr_{N}F^{\dag} 
    + F^\ast r_{\overline{N}}F^{T})_{\alpha \alpha} 
    - \langle \gamma_{N}^{(1)} \rangle 
    (FF^{\dag})_{\alpha \alpha}
    \Bigr] \Biggr\} \,,
\end{align}
where 
\begin{align}
  & \mu \equiv \text{diag}\left[ - \sum_{\beta} C_{\alpha\beta} \, \mu_{X_{\beta}} \right]\,,\\
  &\langle H \rangle 
    = \frac{M_N^2}{2 \, k} 
    + \frac{T^2}{8 \, k} \, F^\dagger F \,,~~~~~
    k = \frac{18 \, \zeta(3)}{ \pi^2 } \, T \,,
  \\
  &\langle \gamma_{N}^{(i)} \rangle 
    = A_{i} \left[ c_{LPM}^{(i)} + y_{t}^{2}c_{Q}^{(i)} 
    + (3g^{2} + g'^{2}) \left( c_{V}^{(i)} 
    + {\rm log} \left( \frac{1}{3g^{2} + g'^{2}} \right
    ) \right) \right] \,,
  \\
  &A_{0} = 2A_{1} = -4A_{2} 
    \equiv \frac{4\pi^{2}}{3 \, \zeta(3)}\frac{T}{3072\pi} \,,
\end{align}
and the values of constants are
$(c_{LPM}^{(0)},\,c_{LPM}^{(1)},\,c_{LPM}^{(2)})=
(4.22,\,3.56, \,4.77)$\footnote{We use the values at $T=10^4~{\rm GeV}$ for simplicity. See the detail in Ref.~\cite{Ghisoiu:2014ena}.},
$(c_Q^{(0)},\,c_Q^{(1)},\,c_Q^{(2)})=
(2.52,\,3.10, \,2.27)$, 
and 
$(c_V^{(0)},\,c_V^{(1)},\,c_V^{(2)})=
(3.17,\,3.83, \,2.89)$.
In the above equations, we use the running coupling constants
at the scale $T$ evaluated from the one-loop renormalization 
group equations.  We use the coupling constants 
at top mass scale as $y_t=0.93690$,
$g' = 0.35830$, $g=0.64779$, and $g_3$ = 1.1666~\cite{Buttazzo:2013uya}.

We solve numerically the kinetic equations
(\ref{eq:KE1})--(\ref{eq:KE3}) and evaluate the chemical potential
$\mu_{X_\alpha}$ at $T=T_{\rm sph}$ where $T_{\rm sph}$
is the freeze-out temperature of the sphaleron process.  In this
analysis we take $T_{\rm sph} = 130$~GeV~\cite{DOnofrio:2014rug}. 
The prediction of the present BAU
is then given by
\begin{eqnarray}
\label{eq:ybmu}
  Y_{B} = 1.3 \times 10^{-3} \sum_{\alpha} 
  \mu_{B/3-L_{\alpha}} (T_{\rm sph}) \,,
\end{eqnarray}
where we have assumed that there is no additional entropy
production for $T < T_{\rm sph}$.
The yield of the BAU depends on 
$M_N$, $\Delta M$, and 
the neutrino Yukawa coupling constants.
Especially, $M_N$ and $\Delta M$ are important parameters
to determine the temperature $T_{\rm osc}$ when their flavor oscillation starts, which is given by
\begin{align}
  \label{eq:Tosc}
  T_{\rm osc} =
  \left( \frac{1}{6} \, M_0 \, \Delta M \, M_N \right)^{1/3}
  =
  320~\mbox{GeV} \,
  \left( \frac{M_N}{3\GeV} \right)^{1/3}
  \left( \frac{\Delta M}{10^{-10}\GeV} \right)^{1/3} \,,
\end{align}
where $M_0 = 7.1 \times 10^{17}~{\rm GeV}$.
In addition, the parameter $X_\omega$ in the neutrino Yukawa 
coupling constants (see Eq.~(\ref{eq:Xom}) in App.~\ref{sec:ap_Yukawa})
is another important parameter for baryogenesis.
This is because it determines the overall scale 
of the Yukawa coupling constants (see, for example,
the discussion in Ref.~\cite{Asaka:2011pb}),
and then the yield of the BAU depends on it significantly.

Now we are at the position to show how the yield of the BAU depends on the
initial value of the matrices of densities.  
We shall discuss the impact of each element separately
and then consider the three cases in which 
we assume the non-zero initial value only for 
(i) $[r_N]_{22}^I$, (ii) Re$[r_N]_{23}^I$, and (iii) Im$[r_N]_{23}^I$.  
We should mention here that $[r_N]_{32}$ also has the initial value satisfying $r_N = r_N^\dagger$ for the cases (ii) and (iii). 
The impact of the initial value $[r_N]_{33}^I$ 
is similar to the case (i), namely the magnitude of the additional contribution is the same but
the sign is opposite.
We should note that the initial value for anti-neutrinos
is taken as $[r_{\overline N}]_{IJ}^I = ([r_{N}]_{IJ}^I)^\ast$
so that there is no asymmetry for any number initially.

\begin{figure}[t]
  \vspace{-0.5cm}
  \centerline{
  \includegraphics[width=10cm]{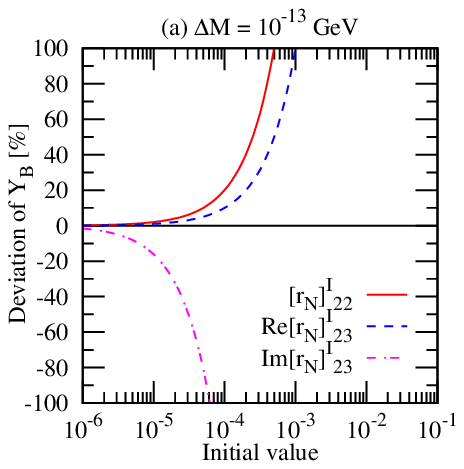}
  \hspace{-4cm}
  \includegraphics[width=10cm]{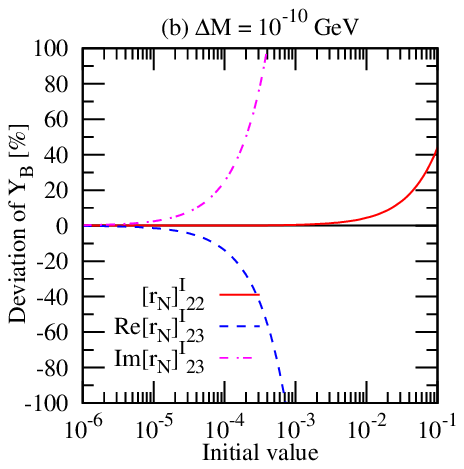}
  \hspace{-4cm}
  \includegraphics[width=10cm]{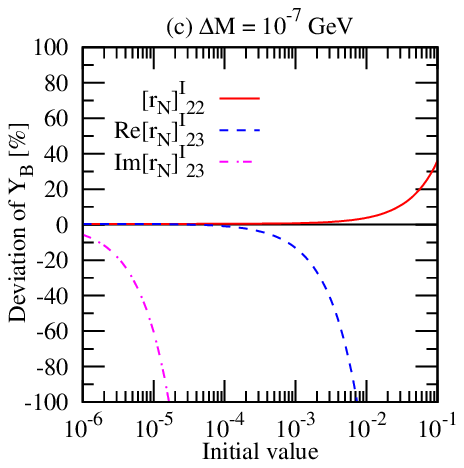}
  }%
  \vspace{-0.5cm}
  \caption{
    Deviation of the yield of the BAU in terms of the initial value
    of the nonzero element of the matrix of densities 
    $[r_N]_{22}^I$ (red solid line), ${\rm Re}[r_N]_{23}^I$ (blue dashed line), 
    and ${\rm Im} [r_N]_{23}^I$ (magenta dash dotted line).
    We fix $X_\omega =1$ and take
    (a) $\Delta M = 10^{-13}$~GeV,
    (b) $\Delta M = 10^{-10}$~GeV,
    and
    (c) $\Delta M = 10^{-7}$~GeV.
  }
  \label{fig:DEV}
\end{figure}
The effect of the non-zero initial value
is represented by the deviation which is defined by
\begin{align}
  \label{eq:dYB}
  \delta Y_B =
  \frac{ 
  \left. Y_B \right|_{[r_N]_{IJ}^I \neq 0} -
  \left. Y_B \right|_{[r_N]_{IJ}^I = 0}   }
  { \left. Y_B \right|_{[r_N]_{IJ}^I = 0}   } \,.
\end{align}
Here $\left. Y_B \right|_{[r_N]_{IJ}^I = 0}$ 
is the yield of the BAU when all initial values are taken to be zero
and $\left. Y_B \right|_{[r_N]_{IJ}^I \neq 0}$
is the one when the initial value is present. 
We fix $M_N = 3$~GeV as a representative value
and take the parameters of the Yukawa coupling constants
explained in Apps.~\ref{sec:ap_Yukawa} and~\ref{Sec:app3}.
We then study how the deviation $\delta Y_B$ 
changes by the initial value of the matrices of densities.

First of all, let us consider the case when $X_\omega = 1$ and the
Yukawa coupling constants are sufficiently small so that the washout
effect is negligible.  
The results of the deviation (\ref{eq:dYB}) are represented in Fig.~\ref{fig:DEV}.
In this case the impact of the initial values diverge 
depending on $\Delta M$.  

For the region with $\Delta M \ll 10^{-10}$~GeV, the oscillation temperature
$T_{\rm osc}$~(\ref{eq:Tosc}) is much smaller than $T_{\rm sph}$
and the oscillation effect to generate the BAU is ineffective.
In this case, as shown in Fig.~\ref{fig:DEV}~(a),
the initial value of each element $[r_N]_{IJ}^I$ 
modifies significantly the prediction of the BAU.
It is seen that the deviation of $Y_B$ becomes larger than 
10~\% when the initial value is large than ${\cal O}(10^{-5})$ in each element.
This critical value from the BAU is much smaller than ${\cal O}(10^{-2})$
which is needed to avoid the overclosure of the universe by $N_1$ with $M_1 \simeq 10$~keV
given in Eq.~(\ref{eq:OMN1}).
To make this point clearer, we also show in Fig.~\ref{fig:FIG_DM_2} the upper bound on $T_R$ 
by requiring that the deviation of $Y_B$ is sufficiently small as $|\delta Y_B| <$ 10~\%. 
It is found that the bound from $\delta Y_B$ is more stringent 
than that from the dark matter density if $M_1 < {\cal O}(10)$ MeV. 
The bound gives the condition for the inflation under 
which the present BAU can be predicted only by the weak scale physics. 
This clearly shows that the prediction of the baryon asymmetry 
can be affected by the processes at the reheating epoch 
even if one demands that such a effect on dark matter is negligible, 
which is opposite to the previous one~\cite{Bezrukov:2008ut}.

\begin{figure}[t]
  \vspace{-0.5cm}
  \centerline{
  \includegraphics[width=10cm]{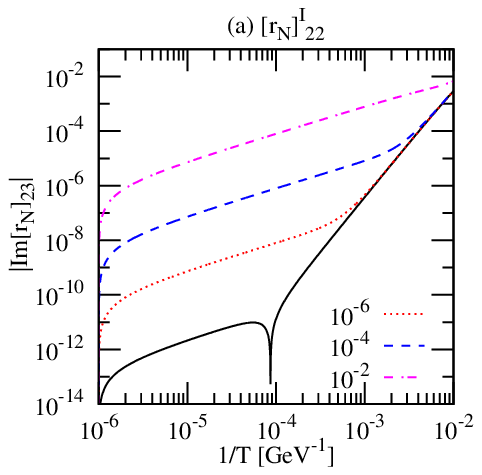}
  \hspace{-4cm}
  \includegraphics[width=10cm]{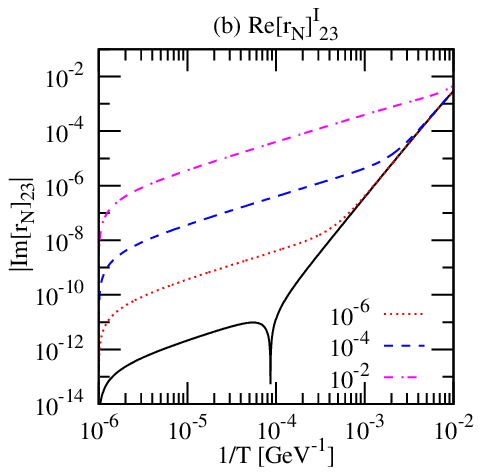}
  \hspace{-4cm}
  \includegraphics[width=10cm]{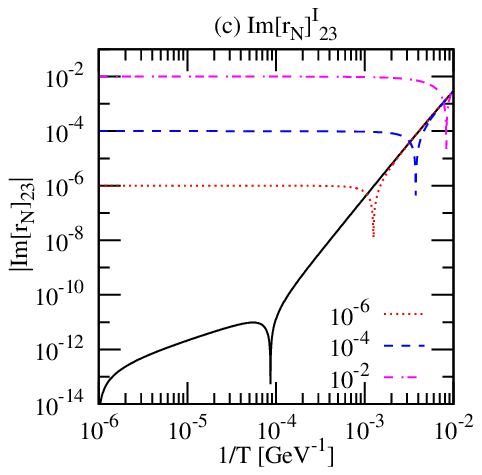}
  }%
  \vspace{-0.5cm}
  \caption{
    Evolution of $\mbox{Im}[r_N]^I_{23}$
    in the presence of the initial value of
    $[r_N]_{22}$ in the panel (a),
    $\mbox{Re}[r_N]_{23}$ in the panel (b), and $\mbox{Im}[r_N]_{23}$
    in the panel (c).
    The initial value is taken as
    $0$ (black solid line),  
    $10^{-6}$ (red dotted line),
    $10^{-4}$ (blue dashed line), and 
    $10^{-2}$ (magenta dot dashed line).
    We take here 
    $\Delta M = 10^{-13}$~GeV, $X_\omega = 1$,
    $\omega_r = + 0.909$, and $\eta = 2.15$.
  }
  \label{Fig:Evo_Imrn23}
\end{figure}

We find that the key parameter in this case is Im$[r_N]_{23}$. 
Each initial value, as shown in Fig.~\ref{Fig:Evo_Imrn23},
gives the additional contribution to Im$[r_N]_{23}$
and, if it alters the value at $T = T_{\rm sph}$ in which
the baryogenesis is most effective, 
the yield of the BAU is modified significantly.
See also Fig.~\ref{Fig:apEvo1} in App.~\ref{sec:C11}.

Next, we consider the region $\Delta M = 10^{-10}$~GeV in which 
$T_{\rm osc} \sim T_{\rm sph}$.
In fact, as shown in Ref.~\cite{Asaka:2005pn},  
this case gives the maximal value of the BAU as a function of
$\Delta M$ (if one can neglect the washout effect).
In this case, as shown in Fig.~\ref{fig:DEV}~(b),
although the initial value of the diagonal element 
does not change the yield of the BAU as long as $[r_N]_{22}^I < \mathcal{O}(10^{-2})$, 
the off-diagonal elements, Re$[r_N]^{I}_{23}$ and Im$[r_N]^{I}_{23}$, 
change the prediction if they become larger than ${\cal O}(10^{-5})$.
The behaviors associated with $[r_N]_{22}^I$ and ${\rm Re} [r_N]_{23}^I$ do not change much even for 
$\Delta M \gg 10^{-10}$~GeV and $T_{\rm osc} \gg T_{\rm sph}$,
but the final BAU becomes more sensitive to Im$[r_N]^{I}_{23}$.
See Fig.~\ref{fig:DEV}~(c).
Therefore, we have found that the initial values of the matrices
of densities can change the prediction of the BAU drastically
even if the overproduction of dark matter is avoided.
Especially, the impact from the off-diagonal elements
is found to be significant in any cases.

\begin{figure}[t]
  \vspace{-0.5cm}
  \centerline{
  \includegraphics[width=10cm]{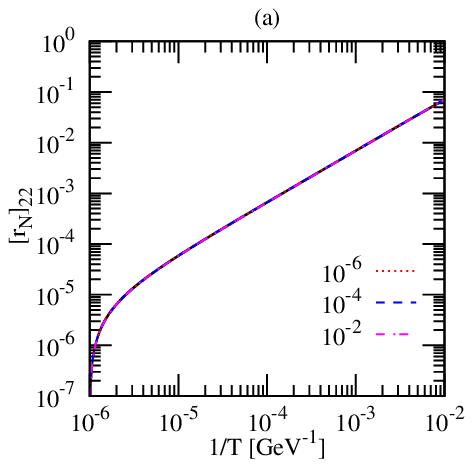}
  \hspace{-3cm}
  \includegraphics[width=10cm]{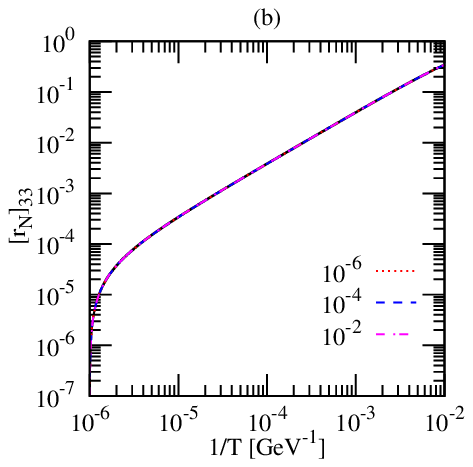}
  }%
  \vspace{-0.5cm}
  \centerline{
  \includegraphics[width=10cm]{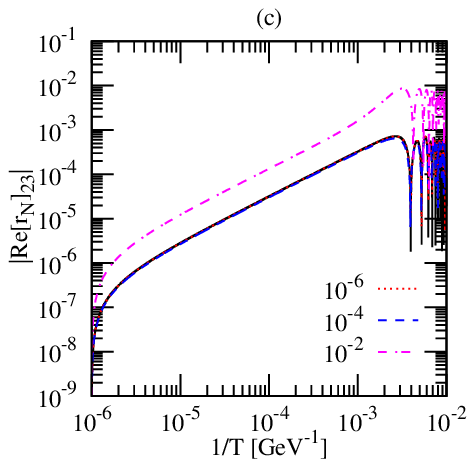}
  \hspace{-3cm}
  \includegraphics[width=10cm]{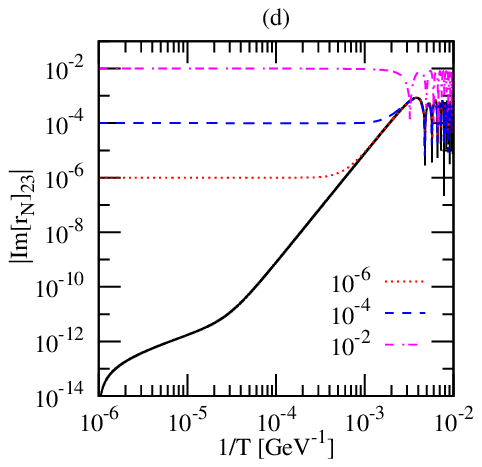}
  }%
  \vspace{-0.5cm}
  \caption{
    Evolution of element of the matrix of densities
    in the presence of the initial value of $\mbox{Im}[r_N]^I_{23}$
    $=0$ (black solid line), 
    $10^{-2}$ (magenta dot dashed line),
    $10^{-4}$ (blue dashed line), and 
    $10^{-6}$ (red dotted line).
    Evolution of $[r_N]_{22}$,
    $[r_N]_{33}$, $\mbox{Re}[r_N]_{23}$, and $\mbox{Im}[r_N]_{23}$
    in the panels (a), (b), (c), and (d), respectively.
    We take here 
    $\Delta M = 10^{-10}$~GeV, $X_\omega = 1$,
    $\omega_r = + 0.01$, and $\eta = 2.15$.
  }
  \label{Fig:Evo1}
\end{figure}
We then discuss the impact of the initial value $\mbox{Im}[r_N]^I_{23}$
in detail.  We show in Fig.~\ref{Fig:Evo1} the evolution of the elements 
of the matrix of densities in the presence of the nonzero $\mbox{Im}[r_N]^I_{23}$.  
It is seen that the diagonal elements $[r_N]_{22}$ and 
$[r_N]_{33}$ do not change much by such an initial value.
In addition, the real part of the off-diagonal element
becomes modified only if $\mbox{Im}[r_N]^I_{23}$
becomes sufficiently large as ${\cal O}(10^{-2})$.
On the other hand, as shown in Fig.~\ref{Fig:Evo1}~(d),
the initial value modifies drastically 
the initial evolution of $\mbox{Im}[r_N]_{23}$, which 
leads to the change of the yield of the BAU.

\begin{figure}[t]
  \centerline{
  \includegraphics[width=10cm]{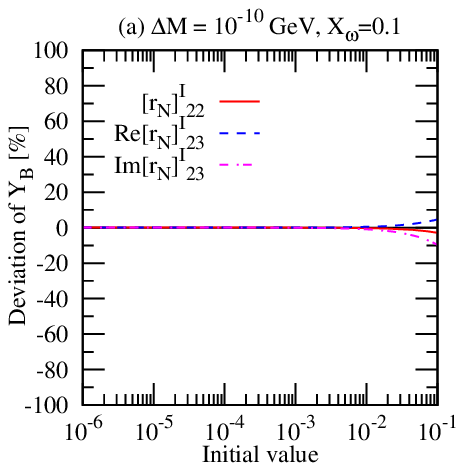}
  \hspace{-3cm}
  \includegraphics[width=10cm]{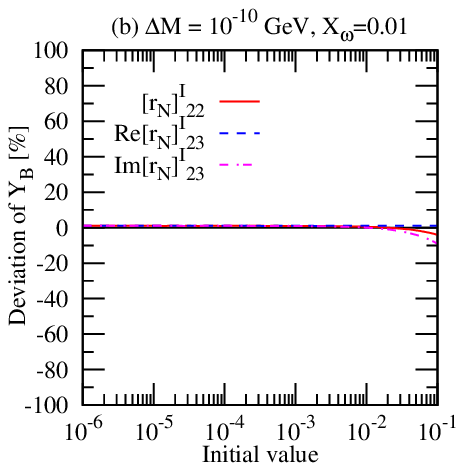}
  }%
  \caption{
   Same as Fig.~\ref{fig:DEV} except for 
    $\Delta M = 10^{-10}$~GeV and taken
    (a) $X_\omega = 0.1$ and 
    (b) $X_\omega = 0.01$. 
  }
  \label{fig:DEV2}
\end{figure}
Finally, we consider the cases when $X_\omega = 0.1$ and $0.01$ where the
overall scale of Yukawa coupling constants are enhanced by a factor 10
and 100 compared with the case $X_\omega = 1$ and then the strong
washout of the BAU occurs.  The deviations of the BAU for these cases are shown 
in Figs.~\ref{fig:DEV2} (a) and (b).  It is interesting to note 
that the initial values of the matrices of densities give no impact on
the yield of the BAU in these cases as long as they are smaller than ${\cal O}(10^{-2})$.
This is expected since the information of the reheating epoch is lost
due to the rapid thermalization processes.
We also find by numerical calculations that 
this behavior does not change when the mass difference becomes 
smaller so that $T_{\rm osc} \ll T_{\rm sph}$ (which should be 
contrasted with the $X_\omega = 1$ case).
Therefore, in the strong washout region the prediction of the BAU
is insensitive to the initial values of the matrices of densities
and the baryogenesis is just the low energy physics
at $T \sim T_{\rm sph}$.

We have discussed the impact of the initial values of the matrices of densities  
on the yield of the BAU $Y_B$ by baryogenesis via neutrino oscillation. 
In addition, we would like to stress here that the CP violation induced by the higher-dimensional operator Eq.~(\ref{eq:HD}) 
is important for the baryogenesis.

\begin{figure}[t]
  \centerline{
  \includegraphics[width=10cm]{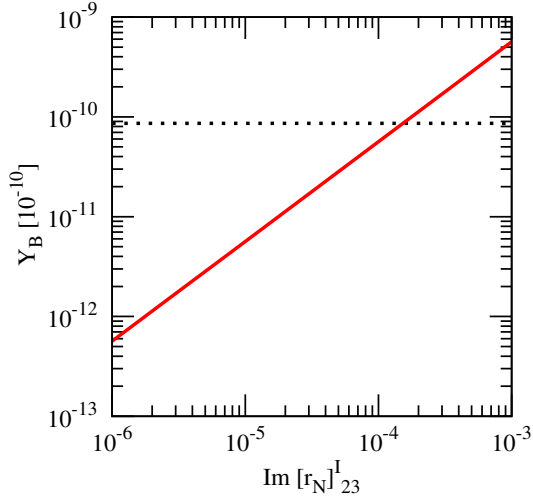}
  }%
  \caption{
  	Yield of the BAU $Y_B$ in terms of the initial value $\mbox{Im} [r_N]^I_{23}$ is shown by the red solid line. 
  	The horizontal dotted line shows the observed BAU $Y_B^{\rm obs}$.
	Here all of the CP violating parameters 
	in the neutrino Yukawa interactions are switched off 
	($\delta = \eta = {\rm Im} \omega=0$). 
  }
  \label{fig:r23-YB}
\end{figure}

In oder to illustrate our point, we consider the case when $\delta=0$, $\eta=0$, and ${\rm Im}\omega=0$, 
i.e. the CP violating parameters in the neutrino Yukawa coupling constants are  turned off all. 
As shown in Fig.~\ref{fig:r23-YB}, the sufficient amount of the BAU can be generated 
by the initial value of $\mbox{Im}[r_N]^I_{23}$ alone.
As an example, the observed value of the BAU is available for $\mbox{Im}[r_N]^I_{23} \simeq 10^{-4}$
when ${\rm Re} \omega = 0.1$, $M_N = 3~{\rm GeV}$, and $\Delta M = 10^{-13}~{\rm GeV}$.
This is a new scenario for the successful baryogenesis.
Notice that we have checked that the sign of $Y_B$ is changed
according to the sign of $\mbox{Im}[r_N]^I_{23}$.
Eventually, we have found that the CP violation in the higher-dimensional
operator as well as that in the neutrino Yukawa interaction can be a source of the BAU.

\section{Conclusions}\label{Sec:conclusion}
In this paper we have investigated the impacts of the higher-dimensional operator 
(\ref{eq:HD}) on the phenomenology of
right-handed neutrinos in the $\nu$MSM.  Especially, we have studied
how the prediction of baryogenesis is altered by the initial
abundance of right-handed neutrinos produced by the scatterings of Higgs bosons 
at the reheating epoch.  We have parametrized such an 
abundance as the initial values of the matrices of densities
$[r_N]^I_{IJ}$ and $[r_{\overline N}]^I_{IJ}$.  Since the coupling matrix
$A$ in Eq.~(\ref{eq:HD}) is nondiagonal generally, the off-diagonal elements of
$[r_N]^I_{IJ}$ and $[r_{\overline N}]^I_{IJ}$ can be present.

The initial abundance of $N_1$ at the reheating epoch may be
responsible for the dark matter density of the present universe, and
the viable parameter space of the $N_1$ mass and couplings is also enlarged. 
In such a case, right-handed neutrino $N_1$ can be dark matter even if
its mass is much large than 10 keV scale, as long as the
stability is ensured due to the discrete symmetry.
It should, however, be noted that these significant effects can be induced 
only if the reheating temperature is very high and/or the cutoff scale is 
lower than the Planck scale.  Otherwise, the effects of the higher-dimensional operator 
is negligible and the dark matter production 
must rely on a low energy physics.

We have also discussed the impact on baryogenesis.  It has been shown
that the initial values do not change the yield of the BAU in the strong
washout region as long as the overproduction of dark matter $N_1$ is
avoided.  Thus, in these cases the baryogenesis in the $\nu$MSM is
insensitive to the physics at high energy scales such as the processes
at the reheating epoch after the inflationary universe.  In this case, the
baryon asymmetry is a consequence of the mechanism at the weak scale
physics.  This strong washout region is very motivated 
since $N_2$ and $N_3$ are expected to be explored 
by future experiments as SHiP~\cite{Anelli:2015pba}, 
LBNE (DUNE)~\cite{Adams:2013qkq}, and FCC-ee~\cite{Blondel:2014bra}
due to the larger mixing elements.

On the other hand, we have found the significant deviation 
due to the initial values in the weak washout region,
which is contrasted to the previous study in Ref.~\cite{Bezrukov:2008ut}.
When the oscillation effect is less important for baryogenesis
($T_{\rm osc} \ll T_{\rm sph}$), the initial values of all the components
are important.  On the other hand, when $T_{\rm osc} \gtrsim T_{\rm sph}$
and the baryogenesis is induced by the oscillation of $N_2$ 
and $N_3$, the initial values of the real and imaginary parts of the off-diagonal element are found to be significant. 
Interestingly, even if all of the leptonic CP violating phases 
contained in the Yukawa coupling constants are switched off ($\delta = \eta = {\rm Im} \omega=0$), 
we have found that the observed amount of the BAU can be obtained 
by the CP violation of $[r_N]_{23}^I$ which is introduced by the higher-dimensional operator alone.
Under the choice of parameters in this analysis, 
any initial value smaller than ${\cal O}(10^{-5})$ is negligible for the prediction of the BAU.
Such a small initial value can be 
obtained if the reheating temperature becomes smaller 
or the cutoff scale becomes higher.
Otherwise, in the weak washout region 
the baryogenesis receives the influence of high energy physics.
In some cases the generation of the BAU can be boosted and hence
the region for the successful baryogenesis in the $\nu$MSM
is enlarged.  

Again, we would like stress that these impacts of the higher-dimensional operator 
are available only if the reheating temperature is very high and/or the cutoff scale is
lower than the Planck scale.  
Otherwise, the baryogenesis in the $\nu$MSM must be explained by the renormalizable Lagrangian.

\section*{Acknowledgments}
The work of T.A. was partially supported by Japan Society for the Promotion of Science (JSPS) KAKENHI
Grant Numbers 15H01031 and 25400249. 
The work of S.E. was partially supported by the Swiss National Science Foundation 200020$\_$162927/1.
T.A., K.M., and T.Y. thank the Yukawa Institute for Theoretical Physics at Kyoto University, 
where this work was initiated during the YITP-S-16-04 on ``the 21th Niigata-Yamagata joint school''. 

\appendix
\section{Production of right-handed neutrinos}
\label{sec:ap1}
In this appendix we discuss the production of right-handed neutrinos
by the operator (\ref{eq:HD}).  To make our argument clear, let us
move to the basis in which the coupling matrix $A$ is diagonal.  This
can be done by the unitary transformation as
\begin{align}
  \label{eq:VR}
  \nu_R' = V_R \, \nu_R \,,
\end{align}
where the mixing matrix $V_R$ is defined by the diagonalization of 
$A$ as
\begin{align}
  V_R^\ast A V_R^\dagger = A'
  = \mbox{diag} (A_1', A_2', A_3' ) \,.
\end{align}
As we will see below, the production is most effective at the
reheating epoch and the basis with $\nu_R'$ is more relevant to
describe the production since the Majorana masses are negligible for
$T_R \gg M_I$.

Let us then consider a pair production of $\nu_{RI}'$:
\begin{align}
  \label{eq:Process}
  \phi_a + \phi_a \to 
  \nu_{R I}' (\vec p_1 \,,~ h_1) 
  + \nu_{R I}' (\vec p_2 \,,~h_2) \,.
\end{align}
where $\vec p_{1,2}$ and $h_{1,2}$ are 
three momenta and helicities of final particles, and
four real scalar fields of the Higgs doublet $\Phi$
are denoted by $\phi_a$ ($a=1,2,3,4$).
The helicity amplitude is given by
\begin{align}
  M_{h_1,h_2} = 
  \frac{1}{\Lambda}
  \bar u (\vec p_1, h_1) 
  \left[ A_I' P_R + A_I'{}^\ast P_L \right] 
  v (\vec p_2,h_2) \,,
\end{align}
where $P_{L,R}$ are chiral projection operators
$P_L = (1-\gamma_5)/2$ and $P_R = (1+\gamma_5)/2$.
By neglecting the masses of right-handed neutrinos, we find
the amplitudes in the center-of-mass frame are given by
\begin{align}
  &M_{+-} = M_{-+} = 0 \,,~~~~~
  M_{++} = + \frac{{A'_{I}}^\ast}{\Lambda} e^{- i \varphi_1} \sqrt{s} \,,~~~~~~
  M_{--} = - \frac{A'_{I}}{\Lambda} e^{+ i \varphi_1} \sqrt{s} \,,
\end{align}
where $\vec p_1 = |\vec p_1| 
(\sin \theta_1 \cos \varphi_1 \,,~
\sin \theta_1 \sin \varphi_1 \,,~
\cos \theta_1 )$ and $\sqrt{s}$ is the center-of-mass energy.
It can be seen that the pair of the same helicity state is produced.
The cross-section of the process (\ref{eq:Process}) is~\cite{Bezrukov:2008ut} 
\begin{align}
  \sigma (\phi_a + \phi_a \to \nu_{RI}' + \nu_{RI}')
  = \frac{ |A_I'|^2 }{16 \Lambda^2} \,,
\end{align}
where the sum of helicities in the final state has been taken.
Then, the scatterings of $\phi_a$ at reheating epoch generate
$\nu_{RI}'$ and its yield for $T \ll T_R$ is given by~\cite{Bezrukov:2008ut} 
\begin{align}
  \frac{n_{\nu_{RI}'}+n_{\overline{{\nu}'}_{RI}}}{s} 
  &= \frac{135 \sqrt{10} \zeta^2 (3)}{16 \pi^8 g_\ast^{3/2}}
  \frac{|A_I'|^2 \, M_P \, T_R}{\Lambda^2} 
    =
    3.68 \times 10^{-6}
  \frac{|A_I'|^2 \, M_P \, T_R}{\Lambda^2}  \,.
\end{align}
This leads to the matrices of densities for right-handed neutrinos
and their antiparticles in the $\nu_{R}'$ basis as
\begin{align}
  [r_{N'}]_{IJ} = [r_{\overline{N'}}]_{IJ}
  = 9.44 \times 10^{-4} 
  \frac{ \, M_P \, T_R}{\Lambda^2}  
  |A_I'|^2 \delta_{IJ}\,.
\end{align}
It is then found from (\ref{eq:VR}) that
the matrices of densities in the $\nu_R$ basis are given by
\begin{eqnarray}
  \label{eq:1}
  [r_N]_{IJ} = 9.44 \times 10^{-4} 
  \frac{M_P \, T_R}{\Lambda^2}  
  \left[ A^\dagger A \right]_{IJ} \,,~~~~~
  [r_{\overline N}]_{IJ} = 9.44 \times 10^{-4} 
  \frac{M_P \, T_R}{\Lambda^2}  
  \left[ A^T A^\ast \right]_{IJ} \,.
\end{eqnarray}
We can see that $r_N^\dagger = r_N$ and
$r_{\overline{N}}^\dagger = r_{\overline{N}}$ as expected,
and also that $r_{\overline{N}} = r_N^\ast$ which means that
there is no asymmetry between $\nu_{RI}$ and $\overline{\nu}_{RI}$ induced by the process~(\ref{eq:Process}).

\section{Yukawa coupling constants of $N_2$ and $N_3$}
\label{sec:ap_Yukawa}
In this appendix we show the parametrization of 
the Yukawa coupling constants of $N_2$ and $N_3$.
We follow the notation in Ref.~\cite{Asaka:2011pb}.

Without loss of generality we can parametrize 
$F_{\alpha I}$ ($I=2,3$) as~\cite{Casas:2001sr,Abada:2006ea}%
\footnote{
We set the coupling for dark matter $N_1$ as $F_{\alpha 1}=0$
for simplicity, since they are severely restricted by the cosmological
constraints~\cite{Adhikari:2016bei}.
}
\begin{eqnarray}
  \label{eq:F}
    F = \frac{i}{\vev{\Phi}} \,
    U \, D_\nu^{1/2} \, \Omega \, D_N^{1/2} \,.
\end{eqnarray}
Here $D_\nu = \mbox{diag}(m_1, m_2,m_3)$ is the mass matrix 
of active neutrinos and the PMNS mixing matrix~\cite{PMNS} is given by
\begin{eqnarray}
  U = 
  \left( 
    \begin{array}{c c c}
      c_{12} c_{13} &
      s_{12} c_{13} &
      s_{13} e^{- i \delta} 
      \\
      - c_{23} s_{12} - s_{23} c_{12} s_{13} e^{i \delta} &
      c_{23} c_{12} - s_{23} s_{12} s_{13} e^{i \delta} &
      s_{23} c_{13} 
      \\
      s_{23} s_{12} - c_{23} c_{12} s_{13} e^{i \delta} &
      - s_{23} c_{12} - c_{23} s_{12} s_{13} e^{i \delta} &
      c_{23} c_{13}
    \end{array}
  \right)  
  \times
  \mbox{diag} 
  ( 1 \,,~ e^{i \eta} \,,~ 1) \,,
\end{eqnarray}
with $s_{ij} = \sin \theta_{ij}$ and $c_{ij} = \cos \theta_{ij}$.  
$\delta$ is a Dirac phase and $\eta$ is a Majorana phase.

Throughout this work we consider the normal hierarchy
of active neutrino masses:
\begin{eqnarray}
  &&m_3 > m_2 > m_1 =0 \,.
\end{eqnarray}
As for the mixing angles and mass squared differences,
we use the central values in Ref.~\cite{Esteban:2016qun}.
\begin{align}
  \begin{array}{l l l}
    \sin^2 \theta_{12} = 0.306 \,, &
    \sin^2 \theta_{23} = 0.441 \,,&
    \sin^2 \theta_{13} = 0.02166 \,,
    \\
    \Delta m_{21}^2 = 7.50 \times 10^{-5}~\mbox{eV}^2 \,,&
    \Delta m_{31}^2 = 2.524 \times 10^{-3}~\mbox{eV}^2 \,.&
  \end{array}
\end{align}
$D_N = \mbox{diag}(M_2,M_3)$ is the mass matrix of HNLs
and we take $M_2 = M_N - \Delta M/2$ and $M_3 = M_N + \Delta M/2$.
The $3 \times 2$ matrix $\Omega$ is given by
\begin{eqnarray}
  \Omega =
  \left(
    \begin{array}{c c}
      0 & 0 \\
      \cos \omega & - \sin \omega \\
      \xi \sin \omega & \xi \cos \omega
    \end{array}
  \right) \,.
\end{eqnarray}
Here $\omega$ is a complex parameter and 
we parametrize its imaginary part as
\begin{align}
  \label{eq:Xom}
  X_\omega = \exp (\mbox{Im} \omega) \,,
\end{align}
and the real part is denoted as $\omega_r$. 
Note that the physical region of these parameters 
is found, for example, in Ref.~\cite{AEIMY}.
In the present analysis, we take the following values in all cases for simplicity. 
\begin{align}
  M_N = 3~\GeV \,,~~~ \xi = 1 \,,~~~
  \delta = 1.45 \,,~~~ \eta = 2.15 \,.
\end{align}

\section{Evolution of baryon asymmetry}
\label{Sec:app3}
In this appendix we shall show the evolution of the yield 
of the BAU depending on the non-zero initial value of the matrices of densities~\cite{Monogawa}.
Note that the absolute value of $Y_B$ is shown in the figures below 
and $Y_B$ is negative in some regions. 
Here, $Y_B$ is not the present value of the BAU but it is given by
\begin{eqnarray}
Y_B (T) 
= 
1.3 \times 10^{-3}
\sum_\alpha 
\mu_{B/3 - L_\alpha} (T) \,,
\end{eqnarray} 
until $T=100~{\rm GeV}$, which should not be confused with $Y_B$ in Eq.~(\ref{eq:ybmu}).

\subsection{Weak washout regime}
First of all, we consider the region in which the Yukawa coupling constants
are so small that the washout process of the asymmetries is ineffective.
\clearpage
\subsubsection{The case with $T_{\rm osc} \ll T_{\rm sph}$}
\label{sec:C11}
\begin{figure}[h]
  \centerline{
  \includegraphics[width=10cm]{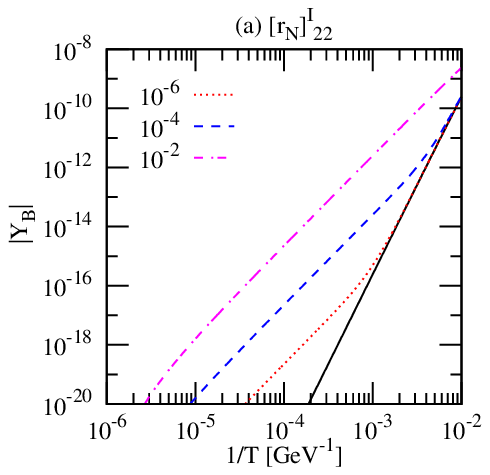}
  \hspace{-3cm}
  \includegraphics[width=10cm]{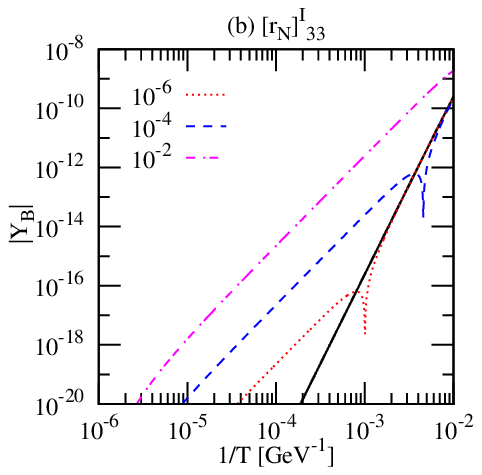}
  }%
  \vspace{-0.5cm}
  \centerline{
  \includegraphics[width=10cm]{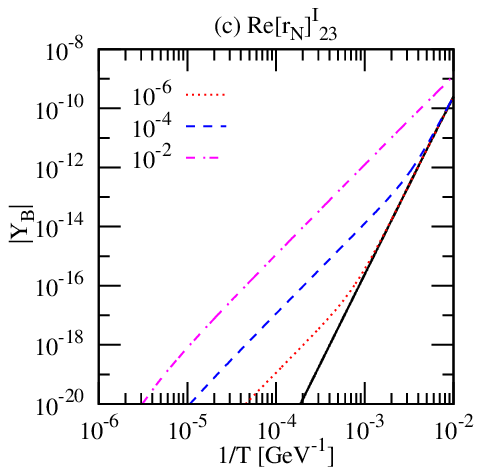}
  \hspace{-3cm}
  \includegraphics[width=10cm]{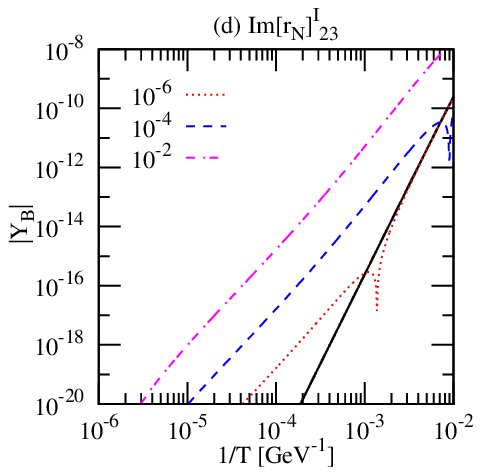}
  }%
  \caption{
    Evolution of the BAU 
    when $\Delta M = 10^{-13}$~GeV and $X_\omega = 1$.
    Black solid line shows the BAU 
    when there is no initial value for all the elements of the matrix of densities.
    The red dotted, blue dashed, 
    and magenta dot dashed lines
    correspond to the cases 
    where the initial value of the element 
    is $10^{-2}$, $10^{-4}$, and $10^{-6}$.  The non-zero initial value is given for $[r_N]_{22}$,
    $[r_N]_{33}$, ${\rm Re [r_N]_{23}}$, and ${\rm Im [r_N]_{23}}$ at the panels 
    (a), (b), (c), and (d).
  }
  \label{Fig:apEvo1}
\end{figure}

We take the following parameter set.
\begin{eqnarray}
  M_N = 3\GeV \,,~~~~~
  \Delta M = 10^{-13} \GeV \,,~~~~~
  X_\omega = 1\,,~~~~~
  \omega_r = + 0.909 \,,~~~~
  \delta = 1.45 \,,~~~~
  \eta = 2.15 \,,~~~~
\end{eqnarray}
which correspond to the case Fig.~\ref{fig:DEV} (a).
The evolution of $|Y_B|$ is shown in Fig.~\ref{Fig:apEvo1}.

\clearpage
\subsubsection{The case with $T_{\rm osc} \sim T_{\rm sph}$}
\begin{figure}[h]
  \centerline{
  \includegraphics[width=10cm]{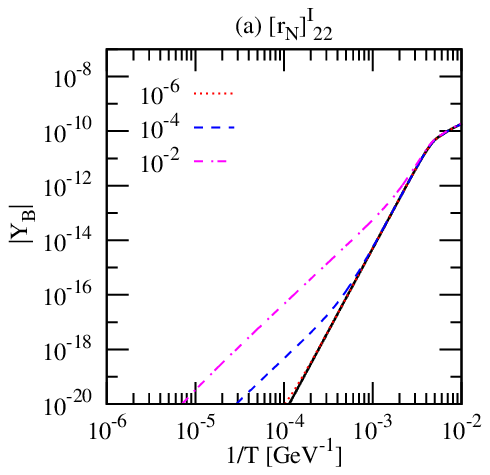}
  \hspace{-3cm}
  \includegraphics[width=10cm]{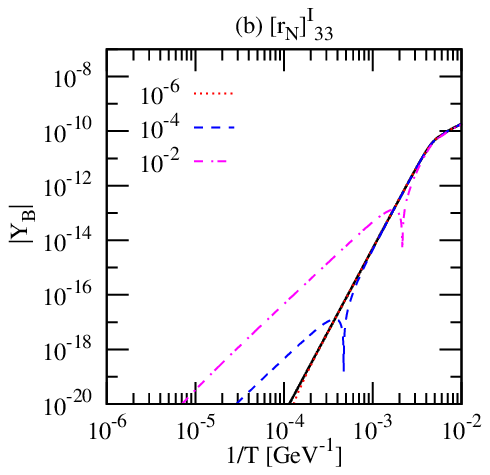}
  }%
  \vspace{-0.5cm}
  \centerline{
  \includegraphics[width=10cm]{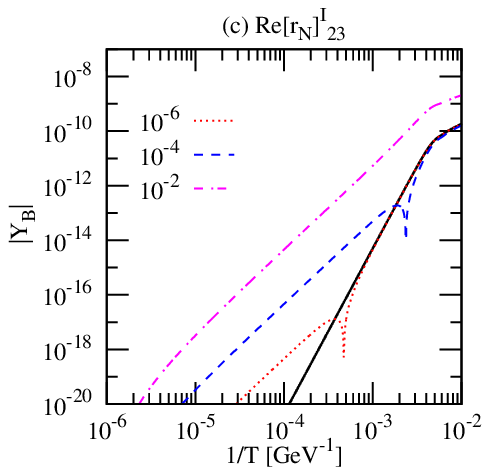}
  \hspace{-3cm}
  \includegraphics[width=10cm]{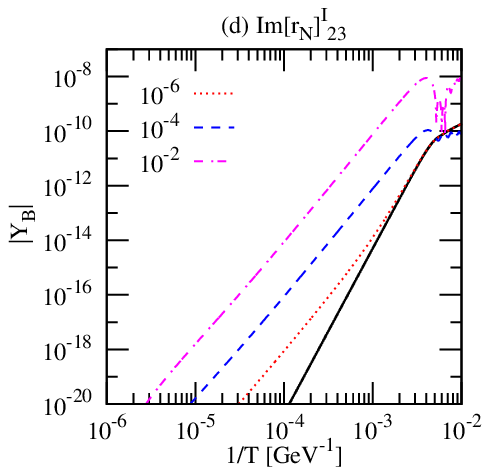}
  }%
  \caption{
   Same as Fig.~\ref{Fig:Evo1} except for $\Delta M = 10^{-10}$~GeV and $X_\omega = 1$.
  }
  \label{Fig:apEvo2}
\end{figure}

We take the following parameter set.
\begin{eqnarray}
  M_N = 3\GeV \,,~~~~~
  \Delta M = 10^{-10} \GeV \,,~~~~~
  X_\omega = 1\,,~~~~~
  \omega_r = + 0.01 \,,~~~~
  \delta = 1.45 \,,~~~~
  \eta = 2.15 \,,~~~~
\end{eqnarray}
which correspond to the case Fig.~\ref{fig:DEV} (b).
The evolution of $|Y_B|$ is shown in Fig.~\ref{Fig:apEvo2}.

\clearpage
\subsubsection{The case with $T_{\rm osc} \gg T_{\rm sph}$}
\begin{figure}[h]
  \centerline{
  \includegraphics[width=10cm]{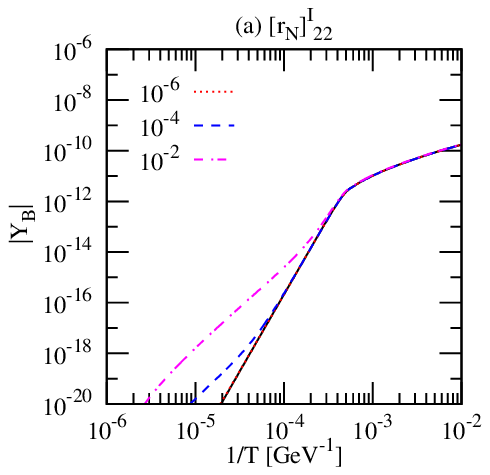}
  \hspace{-3cm}
  \includegraphics[width=10cm]{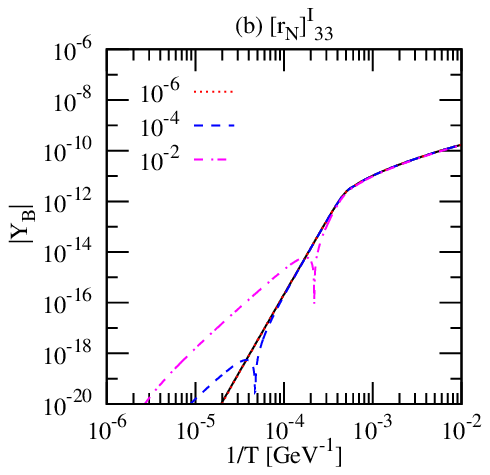}
  }%
  \vspace{-0.5cm}
  \centerline{
  \includegraphics[width=10cm]{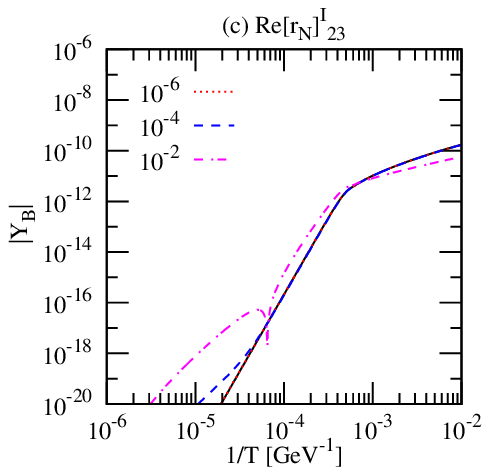}
  \hspace{-3cm}
  \includegraphics[width=10cm]{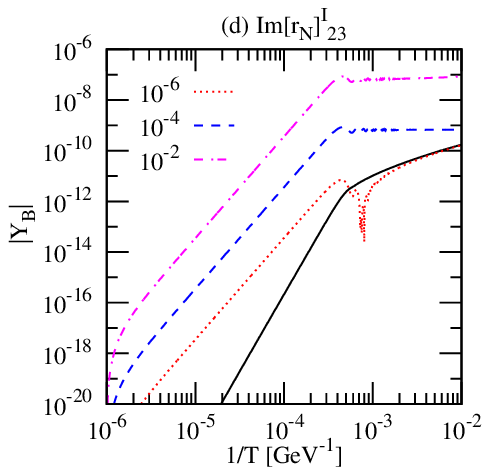}
  }%
  \caption{
   Same as Fig.~\ref{Fig:Evo1} except for $\Delta M = 10^{-7}$~GeV and $X_\omega = 1$.
  }
  \label{Fig:apEvo3}
\end{figure}

We take the following parameter set.
\begin{eqnarray}
  M_N = 3\GeV \,,~~~~~
  \Delta M = 10^{-7} \GeV \,,~~~~~
  X_\omega = 1\,,~~~~~
  \omega_r = + 0.909 \,,~~~~
  \delta = 1.45 \,,~~~~
  \eta = 2.15 \,,~~~~
\end{eqnarray}
which correspond to the case Fig.~\ref{fig:DEV} (c).
The evolution of $|Y_B|$ is shown in Fig.~\ref{Fig:apEvo3}.

\clearpage
\subsection{Strong washout regime}
Next, we turn to consider the region in which the Yukawa coupling constants
are so large that the strong washout of the asymmetries
takes place.

\subsubsection{The case with $T_{\rm osc} \sim T_{\rm sph}$ and $X_\omega=0.1$}
\begin{figure}[h]
  \centerline{
  \includegraphics[width=10cm]{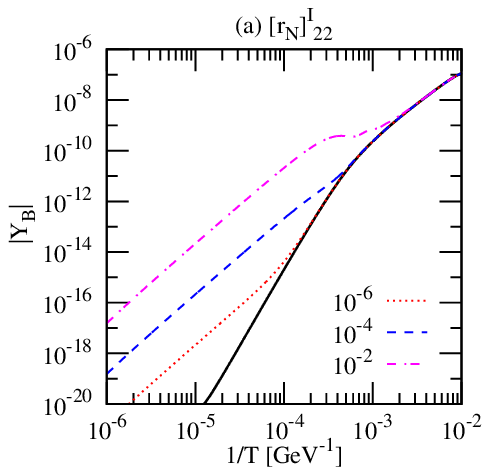}
  \hspace{-3cm}
  \includegraphics[width=10cm]{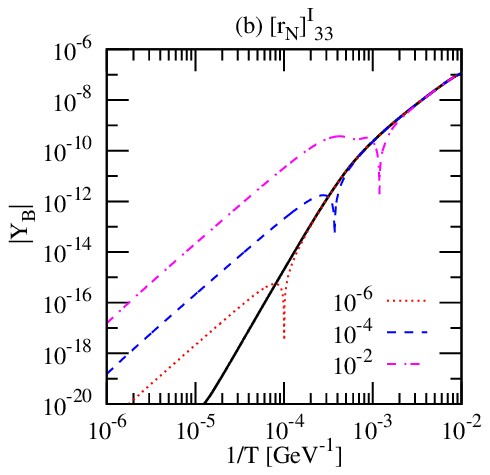}
  }%
  \vspace{-0.5cm}
  \centerline{
  \includegraphics[width=10cm]{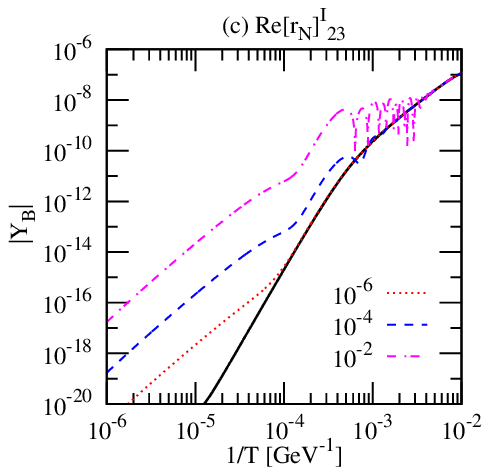}
  \hspace{-3cm}
  \includegraphics[width=10cm]{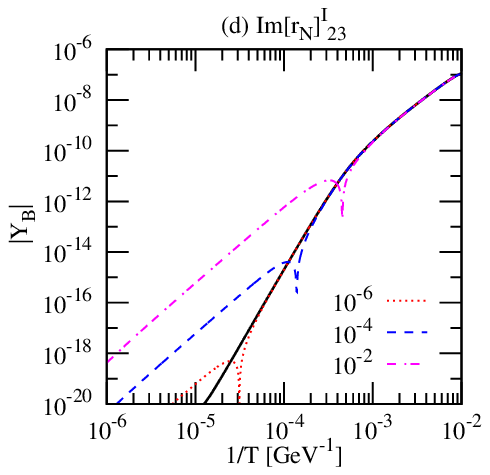}
  }%
  \caption{
   Same as Fig.~\ref{Fig:Evo1} except for $\Delta M = 10^{-10}$~GeV and $X_\omega = 0.1$.
  }
  \label{Fig:apEvo4}
\end{figure}

We take the following parameter set.
\begin{eqnarray}
  M_N = 3\GeV \,,~~~~~
  \Delta M = 10^{-10} \GeV \,,~~~~~
  X_\omega = 0.1\,,~~~~~
  \omega_r = + 0.909 \,,~~~~
  \delta = 1.45 \,,~~~~
  \eta = 2.15 \,,~~~~
\end{eqnarray}
which correspond to the case Fig.~\ref{fig:DEV2} (a).
The evolution of $|Y_B|$ is shown in Fig.~\ref{Fig:apEvo4}.

\subsubsection{The case with $T_{\rm osc} \sim T_{\rm sph}$ and $X_\omega =0.01$}
\begin{figure}[h]
  \centerline{
  \includegraphics[width=10cm]{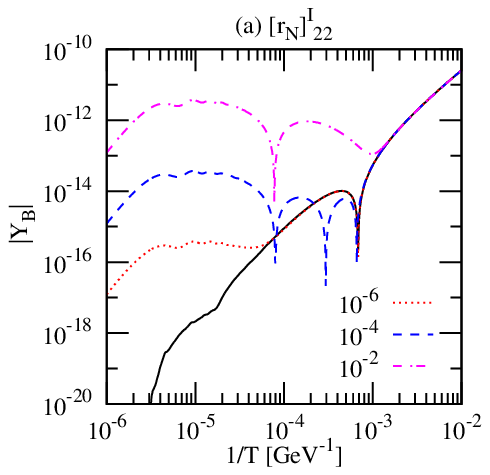}
  \hspace{-3cm}
  \includegraphics[width=10cm]{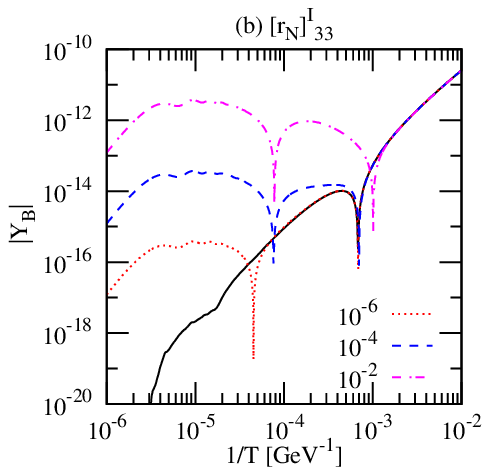}
  }%
  \vspace{-0.5cm}
  \centerline{
  \includegraphics[width=10cm]{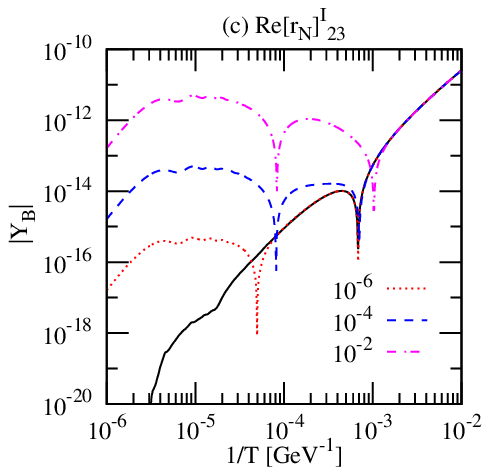}
  \hspace{-3cm}
  \includegraphics[width=10cm]{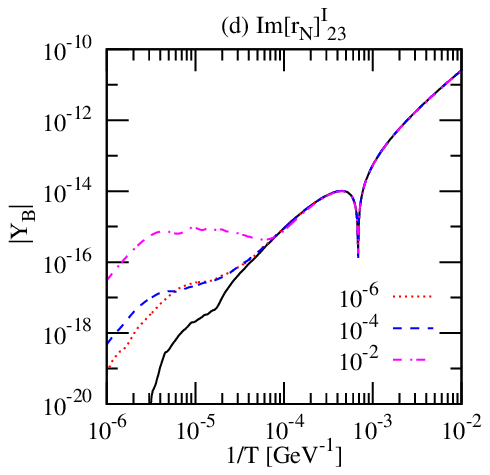}
  }%
  \caption{
   Same as Fig.~\ref{Fig:Evo1} except for $\Delta M = 10^{-10}$~GeV and $X_\omega = 0.01$.
  }
  \label{Fig:apEvo5}
\end{figure}

We take the following parameter set.
\begin{eqnarray}
  M_N = 3\GeV \,,~~~~~
  \Delta M = 10^{-10} \GeV \,,~~~~~
  X_\omega = 0.01\,,~~~~~
  \omega_r = + 0.909 \,,~~~~
  \delta = 1.45 \,,~~~~
  \eta = 2.15 \,,~~~~
\end{eqnarray}
which correspond to the case Fig.~\ref{fig:DEV2} (b).
The evolution of $|Y_B|$ is shown in Fig.~\ref{Fig:apEvo5}.

\end{document}